# Vorticity Confinement and TVD Applied to Wing Tip Vortices for Accurate Drag Prediction

Kristopher C. Pierson[1] and Alex Povitsky[2]
*The University of Akron, Akron, Ohio, 44325-3903*

---

[1] Graduate student, Department of Mechanical Engineering, College of Engineering at the University of Akron, Akron OH 44325-3903, AIAA Student Member
[2] Professor, Department of Mechanical Engineering, College of Engineering at the University of Akron, Akron OH 44325-3903, Associate Fellow AIAA, email: povitsky@uakron.edu, the corresponding author

**Abstract**

**The vorticity confinement (VC) method was used with total variation diminishing (TVD) schemes to reduce possible over-confinement and applied to tip vortices shed by edges of wings in order to predict induced drag using far-field integration. The optimal VC parameter was determined first by application to 2-D vortices and then to tip vortices shed by a 3-D wing. The 3-D inviscid simulations were post-processed using the wake-integral technique to determine lift-induced drag force. Dependence of the VC parameter on the flight Mach number and the angle of attack was evaluated. Grid convergence studies were conducted for 2-D vortices and for induced drag generated by 3-D wing. VC was used with TVD minmod and differentiable flux limiters to evaluate their effect on the VC method. Finally, the VC approach was combined with the Reynolds stress equation turbulence model, and the results were compared to experimental data of tip vortex evolution.**

**Keywords:** Vorticity confinement, Total variation diminishing (TVD) schemes, tip vortex, turbulent flow

**Nomenclature**

| | |
|---|---|
| $c$ | Coefficient of vorticity confinement |
| $C$ | Chord length |
| $F$ | Flux |
| $F_d$ | Drag force |
| $F_{d_i}$ | Induced drag |
| $h$ | Characteristic size of grid cell |
| $K$ | Kinetic energy |
| $\vec{n}$ | Surface normal vector |
| $n_x$ | Number of grid points in the x direction |
| $n_y$ | Number of grid points in the y direction |
| $P$ | Pressure |
| $\vec{q}$ | Vector of momentum |
| $r$ | Radial coordinate |
| $r_0$ | Radius of Taylor vortex |
| $\vec{s}$ | Vorticity confinement source term |
| $t$ | Time |
| $t_{wing}$ | Thickness of wing |
| $u$ | Velocity in the x direction |
| $v$ | Velocity in the y direction |
| $\vec{v}$ | Vector of velocity |
| $V_\theta$ | Tangential Velocity |
| $w$ | Velocity in z direction |
| $\alpha$ | Angle of attack |
| $\Gamma$ | Circulation |
| $\varepsilon$ | Confinement parameter |
| $\mu$ | Dynamic viscosity |
| $\nu$ | Kinematic viscosity |
| $\rho$ | Density |
| $\tau$ | Shear stress |
| $\bar{\bar{\tau}}$ | Viscous stress tensor |
| $\Phi$ | Flux limiter function |
| $\omega$ | Vorticity |

# I. Introduction & Background

The research presented in this study was motivated by the need to achieve reliable prediction of lift-induced drag (or simply, induced drag), which is associated with the formation of tip vortices, and to minimize numerical dissipation of tip vortices. Another goal was to improve accuracy of vorticity confinement approach and avoid possible over-confinement by coupling it with total variation diminishing (TVD) schemes. Induced drag becomes the major component of drag for aircraft in subsonic flight with shorter aspect ratio wings, higher angle of attack, and higher Mach numbers. Utilization of computational fluid dynamics (CFD) for the design of winglets and wing planforms for the purpose of induced drag reduction requires multi-variant design optimization using moderately sized numerical discretization grids that allow for only a few grid points per tip vortex diameter.

Aerodynamic forces from a CFD simulation are most commonly determined by the integration of pressure and shear stress over the surface of the aerodynamic body. This method, known as the near field technique, generally provides accurate lift prediction; however, accurate induced drag prediction can be difficult to obtain barring a sufficiently refined grid at the surface of the aircraft [1]. The loss of accuracy in the prediction of induced drag occurs because it involves numerical integration of pressure in areas of strong pressure gradients near wing tips, where tip vortices form. With the near-field integration method this problem could, in principle, be resolved by increasing surface grid density to better represent curvature of the aircraft. This, however, can be problematic for preliminary aerodynamic design as moderate-sized grids should be used to keep computational costs to a minimum to increase turn-around time. Induced drag simulations by Snyder and Povitsky [2] showed that the near-field method over-predicts (approximately double) the induced drag, even if a reasonably refined grid was employed.

Alternatively, drag can be predicted from a technique known as the far-field or wake-integral technique. This technique applies conservation of mass and momentum to a control volume which encloses the wing. The integration can be simplified under the assumption of 1-D flow in the far-field to include only integration over a cross-flow plane within the wake; this plane is known as the Trefftz plane [3]. Such a method overcomes many of the issues present in the near field technique such as strong pressure gradients since the integration is performed in the far flow field instead of the surface of the aircraft. Vos et al. [1] demonstrated the advantages of far-field drag prediction compared to near-field integration. Their results [1] showed that the spread in drag coefficients obtained on different grids is much lower for drag computed by far-field integration compared to near-field integration. The grid convergence drag characteristics obtained from near-field and far-field analysis are shown in Ref [4], Fig. 8. In this figure, far-field drag has almost constant values from the coarse, medium and fine grids within 2 drag counts. The drag values are close to the expected extrapolated value of the near-field drag convergence study. On the contrary, the near-field drag is grid-dependent and the extrapolated value is not achieved as it requires impractically refined grid. Grid dependence is also investigated in the current work in Sections II B and III B.

Additionally, the far-field integration technique decomposes the components of drag into induced viscous and wave drag components [5]. As shown in Ref [2], to obtain the drag force by the Trefftz plane method the following integrals should be computed:

$$F_D = -\left[\oint P_{in} dS - \oint P_{out} dS + \oint \rho u^2 dS - \oint \rho_\infty U_\infty^2 dS\right] \quad (1)$$

The lift-induced component of drag is given by the integration across the wake in the far field (in the Trefftz plane):

$$F_{D_i} = \int_{wake} \frac{\rho_\infty}{2}(v^2 + w^2) dy dz \quad (2)$$

However, this alternative technique under-estimates induced drag due to numerical dissipation (an unavoidable part of CFD simulations) of the tip vortices as they convect from the wing to the Trefftz plane.

CFD solvers typically employ upwind discretization schemes for convective terms, and while the schemes improve numerical stability, they come at the cost of elevated numerical dissipation compared to central difference schemes. Numerical dissipation causes decay of vortices at a much greater rate than physical dissipation [2]. As the vortices convect downstream, this decay is clearly observed through the rapid decrease of peak velocity and expansion of the vortex core, which leads to an under-prediction in wake-integrated drag. Vorticity confinement (VC) can be used to improve the results of the wake-integral method for prediction of induced drag without the computational cost of a highly refined grid. The VC method can be used to counteract nonphysical numerical dissipation and eliminate the decay of vortices even for coarse grids, which would normally be plagued by unnatural levels of dissipation through numerical viscosity. The VC method can be implemented through the addition of a body force term to the momentum equation [6,7,8,9], as shown in Eq. (3).

$$\frac{\partial \rho \vec{u}}{\partial t} + \vec{u} \cdot \left(\nabla(\rho \vec{u})\right) = -\nabla P + \nabla \cdot \bar{\bar{\tau}} - \rho \varepsilon \vec{s}, \quad (3)$$

where $\bar{\bar{\tau}}$ is the viscous stress tensor [10, 27].

The vector $\vec{s}$ is given by Eq. (4) as the product of the direction vector $\vec{n}$, defined in Eq. (6), and the local pseudo-vector of vorticity, $\vec{\omega}$.

$$\vec{s} = \vec{n} \times \vec{\omega} \tag{4}$$

The confinement parameter, $\varepsilon$, is either set as a constant or can be dynamically determined using Eq. (5).

$$\varepsilon = ch^2 |\nabla|\vec{\omega}|| \tag{5}$$

The constant $\varepsilon$ formulation is the original formulation of VC, and $\varepsilon$ determined by Eq. (5) is a grid and vortex strength dependent formulation of VC, which was originally proposed by Lohner et al. [7]. This methodology is denoted herein as modified VC. The intent of this formulation is to scale the confinement parameter appropriately for use with second order schemes. In Eq. (5), the dimensionless coefficient $c$ is the constant coefficient of confinement parameter and can be used for tuning the strength of the confinement source term. Also, the variable $h$ is the length scale of the local finite volume cell, and $\vec{\omega}$ is the local pseudo-vector of vorticity. The direction of vector $\vec{n}$ is calculated by Eq. (6), and the length scale $h$ is determined by Eq. (7).

$$\vec{n} = \frac{\nabla |\vec{\omega}|}{|\nabla|\vec{\omega}||} \tag{6}$$

$$h = \sqrt{(\Delta x)^2 + (\Delta y)^2 + (\Delta z)^2} \tag{7}$$

In Eq. (7), the values of $\Delta x$, $\Delta y$ and $\Delta z$ are determined by differencing the maximum and minimum nodal position of a particular cell in the each of the three cartesian direction. Another expression for $h$, Eq. (15), is introduced for the structured 3-D grid used in Section IV. Apart from second-order schemes, there was a recent attempt by Hejranfar et al. [11] to combine VC with a high-order solver and to show that using high-order compact finite-difference schemes in conjunction with the vorticity confinement method could reasonably preserve the structure of vortical flows. In their conclusion, the researchers [11] note the importance of the selection of an appropriate confinement parameter, which has the strength to preserve the vortex while also not resulting in non-physical over-confinement or strengthening of the vortex. The modified VC method investigated in this study was developed to properly scale the confinement term so that a single variable, $c$, achieves this outcome for a wider range of flight parameters, geometries and grid densities.

Regarding higher order schemes, work by Costes et al. [12,13] sought to extend another type of VC, typically denoted as VC2 [14], to higher order schemes. The original VC method was more desirable in the current study due to the application to a wider range of problems including non-uniform, unstructured grids [2, 6, 7, 8, 9, 11, 15, 16, 17]. The referenced studies were conducted without objectionable effects arising for the method's non-conservative nature.

In prior research [2], inviscid computations were conducted to have induced drag as the only drag component in order to compare CFD results to aerodynamic lifting line drag prediction in a straightforward way. A study by Destarac [18] was purposefully restricted to solutions of the Euler equations so that induced drag could be analyzed using the far-field approach without dissipation occurring from viscous effects.

In Ref [2], modified VC was applied to improve the calculation of drag using the wake-integral method. It was shown that modified VC makes induced drag computations independent of the Trefftz plane location and much closer to aerodynamic lifting line theory compared to surface integration. It was also shown that VC reduced the level of numerically generated entropy, which may reduce the threshold for the entropy correction method [2].

To summarize Refs. [2, 8], the wake-integral technique was used to process the results of inviscid simulations conducted using the VC method, and the following conclusions were made.

- Wake-integral induced drag prediction was more accurate compared to pressure integration at the wing surface.
- VC combined with the wake-integral technique improved the preservation of trailing vortices and made induced drag computations independent of the Trefftz plane location.
- VC prevented the shift from induced drag to spurious entropy drag.

The VC approach developed in [2] to improve the determination of induced drag for low angle of attack (AoA) (AoA= 4 degrees) and low Mach number ($M=0.3$) was extended in the current study to higher AoAs as well as higher velocity flows (yet still subsonic), and the optimal value of $c$ for counteracting numerical dissipation was determined for each of these cases. This is an area of particular interest as it was concluded in a recent dissertation [19] that "there is still a lot of work that needs to be addressed in order to identify the best value of the confinement parameter for a given case. Discretization errors, numerical diffusion, numerical schemes and geometries all have to be taken into account." In the presented research, the VC approach was also applied to second order numerical discretization coupled with total variation diminishing (TVD) schemes [20]. The TVD schemes were investigated for their capability of preventing over-confinement of vortices. While the comparison of TVD limiters was conducted in literature [20], TVD was not used together with VC, to the best of the authors knowledge.

In addition to wake-integral technique for drag prediction, the behavior of vortices plays a role in determining safe distances between aircraft in high lift conditions such as takeoff and landing, interactions between shed vortices and the aircraft's tail as well as vibrational noise caused by submarine sails. A common goal of tip vortex studies is to reduce the inefficient or hazardous effects trailing vortices have on other lifting surfaces or structures which may lie in their path (see [21] and references therein). VC could be used to improve the accuracy of modeling this type of interaction. In the current study, VC was applied along with FLUENT's implementation of the Reynolds stress turbulence model to simulate the evolution of a wingtip vortex; this simulation was a replica of a physical experiment in which velocity profile measurements were taken. The performance of the VC methodology was then evaluated through the comparison of simulation results to the experimentally measured velocity profiles. This comparison shows that the VC methodology can be used to capture the dynamics of turbulent dissipation of tip vortices by eliminating the unnatural numerical dissipation that occurs even in large cell count 3-D meshes (in the current, study numerical dissipation is much greater than physical dissipation despite more than 10 million grid cells in the mesh).

The study is composed as follows. To select VC parameters, Euler computations using the authors' in-house MATLAB-based code for 2-D Taylor vortices with and without crossflow are presented in Sec. II. In Sec. II B, a combination of VC with various TVD schemes are discussed to show that TVD could suppress over-confinement and related oscillations in the solution and, therefore, make the choice of VC parameter more flexible. The methodology of incorporating VC into the discretized Euler equations developed in Section II was used for computations presented in subsequent Sections III and IV. In Sec. III, the effect of VC is evaluated for subsonic 3-D flow over a finite-span wing for a range of angles of attack (AoA) from 4 to 10 degrees and Mach numbers of 0.3, 0.5 and 0.6. It is shown that the method was accurate for higher subsonic flight speeds and at higher angles of attack. Values of vorticity confinement parameters obtained from simulations presented in Section III were used as a starting point for computations discussed in Section IV in which the VC parameter still required retuning since the freestream velocity, size of the wing and angle of attack were different from the previous section. The values of VC parameters selected for 2-D examples using a uniform grid in Section II are different from those used with 3-D non-uniform grids in the next sections. Reasons for the discrepancy of VC parameters values include non-uniformity of the grid, the fact that second-order of numerical discretization may or may not be obtained on local grid cells (as shown in Section II C) and the non-orthogonal property of local unstructured grids used in Sections III and IV. Additionally, it was found in Section II D that FLUENT produces lower levels of numerical dissipation compared to the authors' code which results in a lower strength VC source term to properly maintain the vortex. Section IV includes effects of turbulence on tip vortices by using the VC method with the Reynolds stress turbulence model. The results of numerical modeling are compared to experimental velocity profiles in shed tip vortices.

## II. Application of Vorticity Confinement to Convection and Dissipation of Vortices in Uniform Flow

### A. Numerical Solution of Euler/Navier-Stokes Equations

For simulation of convection and dissipation of vortices in uniform flow, a compressible Navier-Stokes solver was developed by the authors and had the following properties: structured grid with finite-volume discretization, Steger-Warming flux splitting [20], second-order upwind discretization of convective terms, central discretization of viscous terms, and first-order explicit time stepping. For computation of VC body force, Eqs. (4-7), vorticity and vorticity gradient are computed by a central difference scheme. The addition of the VC terms was found to increase the computational time by 19% for the simulations carried out in this section using MATLAB.

The original VC formulation had the degree of confinement controlled by a heuristic constant called the confinement parameter. A methodology was developed with the goal of creating an adaptive, automatic estimate of the confinement parameter, but it was limited to academic cases of vortex convection in uniform flow [22]. In the present research, both the original VC formulation and a modified VC scheme with a scaled confinement parameter formulation [4,23] (see Eq. (4)) were used. Modified VC features a true unit-less constant $c$ (as opposed to unit variable $\varepsilon$). The $h^2$ multiplier of the VC source term ensures that the strength of confinement scales similar to the leading truncation error term for a second-order scheme [20]. This was a useful property of the scheme given the goal of counteracting numerical dissipation in unstructured grids with varying grid density. The correct choice of constant $c$ for a set of representative cases was one of the goals of the investigation.

The Taylor vortex was used to initiate the velocity of the flow field. The rotational velocity is given as follows:

$$V_\theta = \frac{r}{r_0} \frac{\Gamma}{2\pi} \exp\left(0.5\left(1 - \left(\frac{r}{r_0}\right)^2\right)\right) \tag{8}$$

where $\Gamma$ is vortical velocity circulation and $r_0$ is initial vortex core radius.

The domain was rectangular with a normalized by $r_0$ length of 50 in the $x$ direction (direction of mean flow) and 20 in the $y$-direction with a grid step size of $\Delta x = \Delta y = 0.2$, which allowed for ten grid intervals over the vortex diameter. The goal was to test the vorticity confinement approach on a relatively coarse grid. The total time of the simulation was $t = 0.2$ s consisting of 4000 time steps, the Courant number was 0.025. The vortex convected across the flowfield in the $x$ direction due to $u_{cf}$, a uniform flow in the $x$ direction of $M = 0.3$. For the selected value of convective velocity, the center of vortex moved approximately 20 units from its initial location at the origin by the end of time interval $t = 0.2$. The domain size was selected so as the vortex remains within its boundaries.

The properties of the initial vortex for these simulations were chosen as $r_0 = 1$ and $\Gamma = 50$. For displaying the results, vorticity was normalized by $\bar{\omega} = \frac{\omega}{\Gamma \cdot r_0^2}$ for all vorticity contour plots in this Section. The contour plot's spatial coordinates were normalized by $\bar{x} = \frac{x}{r_0}$ and $\bar{y} = \frac{y}{r_0}$ with $x = 0$ and $y = 0$ at the center of the vortex. For Fig. 1a and Fig. 2a, velocity was normalized by the initial peak velocity of the vortex ($V_{\theta,max} = \frac{\Gamma}{2\pi r_0}$), that is, $\overline{V_\theta} = V_\theta / V_{\theta,max}$. The radial coordinate for the line plots was normalized as $\bar{r} = \frac{r}{r_0}$. Values of velocity at negative $\bar{r}$ were shown to confirm that the vortices are symmetric.

Figure 1a shows the decay of the tangential velocity of the vortex with and without VC through comparison to the initial vortex. A comparison of Figs. 1c and 1d to Fig. 1b shows the same decay through a contour plot of vorticity in the region surrounding the vortex. While the second-order upwind scheme showed improvement in the level of numerical dissipation compared to the first-order upwind scheme simulations conducted by the authors in [24,25], the results still show that the peak velocity and vorticity decay significantly when the numerical simulation runs without VC implemented. Recall that viscous terms were not included here; thus, the dissipative effects were due solely to the numerical scheme. Fig. 1a and Fig. 1d show improvement when VC is used to counteract numerical dissipation and match more closely with the initial vortex vorticity field (Fig. 1b).

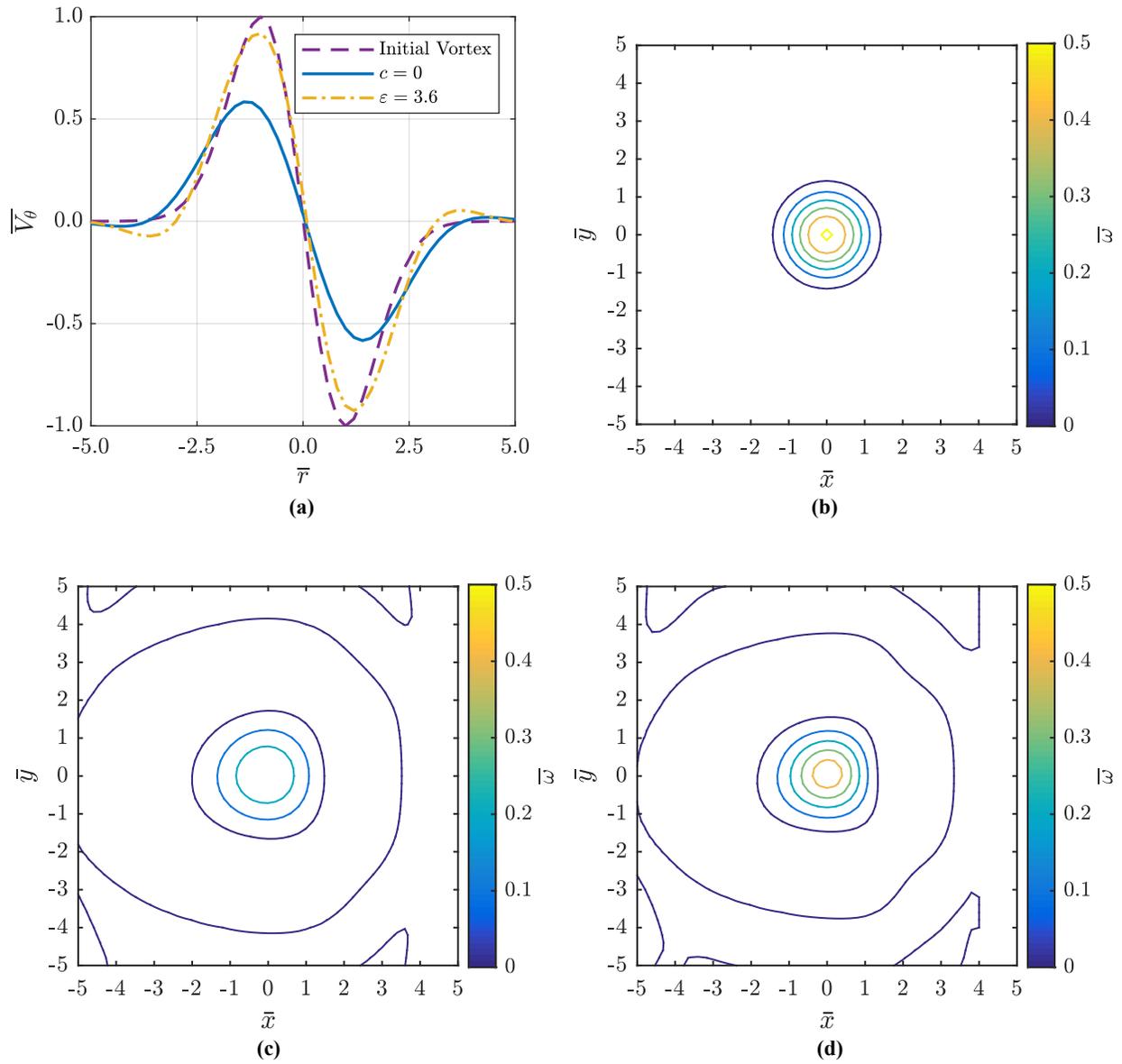

**Fig. 1** Numerical dissipation using upwind 2nd order discretization: (a) normalized tangential velocity without VC and with VC, ε=3.6; (b-d) contours of normalized vorticity: (b) initial profile; (c) without VC; and (d) with VC.

The method was further extended to include the modified VC formulation which utilizes Eq. (5). The optimal coefficient of confinement parameter, $c$, was found to be 2.3. Kinetic energy for the region was determined using Eq. (9).

$$K = \int_0^R \int_0^{2\pi} \frac{1}{2} \rho \left( (u - u_{cf})^2 + v^2 \right) r d\theta dr \tag{9}$$

Fig. 2a is a plot of normalized kinetic energy calculated based on motion relative to the mean flow. Kinetic energy is normalized by $\bar{K} = \frac{K}{K_{init}}$, where $K_{init}$ is the kinetic energy from the initial conditions of the system. Time is normalized in the plot, $\bar{t} = \frac{t \cdot V_{\theta,max}}{2\pi r_0}$, making $\bar{t} = 1$ the time it takes for a particle to rotate around the vortex at $r = r_0$. To eliminate the effect of numerical noise left along the path of the vortex, only the region within $3r_0$ of the vortex center was evaluated. The region that this radius encompassed moved in time and was defined by $R = \sqrt{(x - t \cdot u_{cf})^2 + y^2}$. In addition to the tracking of kinetic energy, momentum was also tracked throughout the simulation. The absolute value of momentum was selected to be certain that the positive and negative values on opposite sides of the vortex were not simply being increased at the same time.

$$M_x = \int_0^R \int_0^{2\pi} \frac{1}{2} \rho |u - u_{cf}| r d\theta dr \tag{10}$$

Figure 2b shows the absolute value of the $x$ momentum determined by Eq. (10) and normalized by its initial value ($\overline{M_x} = \frac{M_x}{M_{x,init}}$). Figures 2a and 2b show that VC with $c = 2.3$ had only a minor effect on integrals of momentum and kinetic energy, with $x$ momentum increasing by a small amount throughout the simulation and kinetic energy decreasing slightly. It is shown in Figs. 2a,b that increase of c from 2.3 to 2.75 causes non-physical increase of kinetic energy and momentum compared to initial physical values of these variables. The optimum value of $c$ can be determined either by a constant level of kinetic energy (Fig. 2a), by tracking the absolute value of momentum (Fig. 2b) or by the best correspondence to initial vortex velocity profile (Fig. 2c). Note that the listed criteria might lead to slightly different values of optimal $c$. For example, increase of $c$ from 2.3 to 2.5 would result in a vortex velocity profile that is closer to its initial profile (compare Fig 2c to Fig. 3a); however, it also would cause further increase of momentum (Fig. 2b) compared to the initial condition.

By observation of Fig. 2c, the maximum vortex velocity was slightly less than its initial value even for $c = 2.3$; therefore, the kinetic energy decreased slightly (Fig. 2a). Kinetic energy is a quadratic function of vortical velocity (see Eq. (9)), which tends to amplify maximums of velocity. As the maximum velocity decreased, the kinetic energy decreased. On the other hand, the numerical vortex became wider by a small amount compared to the initial vortex (see Fig. 2c). This widening caused a slight increase in the absolute value of $x$ momentum (Eq. 10 and Fig. 2b).

Figures 2c,d,e are plots for $c = 2.3$ and respectively show a line plot of the vortex tangential velocity, contour plot vorticity, $\omega$, and a contour plot of confinement parameter, $\varepsilon$. The velocity profile and vorticity plots in Fig. 2c and 2d are qualitatively similar to those in Fig. 1a and 1d. Fig. 2e shows contours of $\varepsilon$ determined by Eq. (5). Comparing Fig. 1 to Fig. 2, it can be seen that there was little difference between the simulation conducted using constant $\varepsilon$ and that using variable $\varepsilon$ determined by Eq. (4). Nevertheless, it was necessary to implement modified VC for 3-D simulations using unstructured non-uniform grid as the size of the finite-volume cell correlates to the amount of numerical dissipation. Modified VC considers the grid size and either strengthens or weakens the level of confinement based on both the strength of the vortex and the local spatial step size.

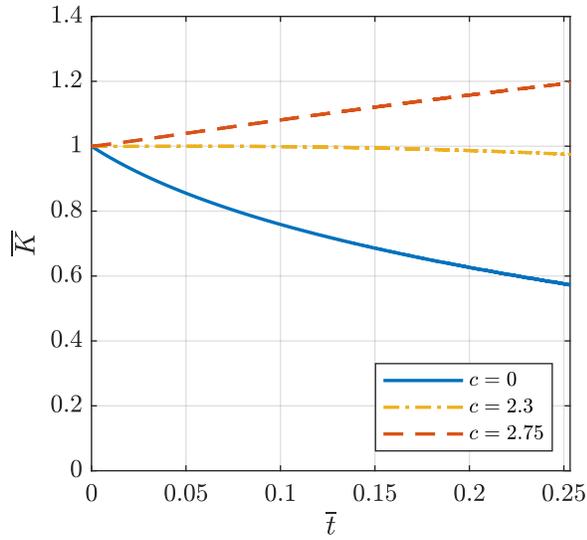

(a)

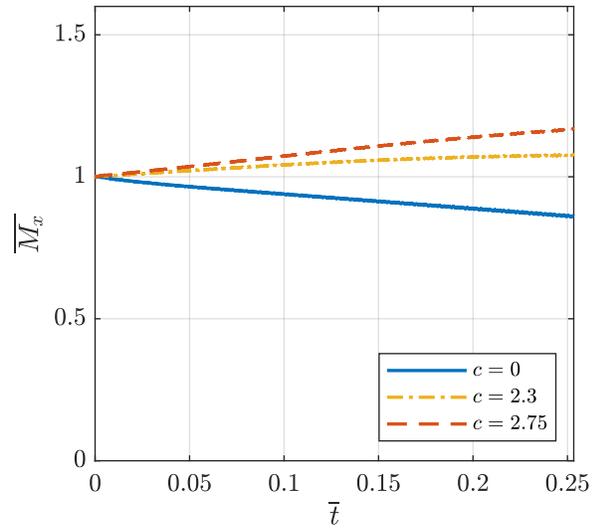

(b)

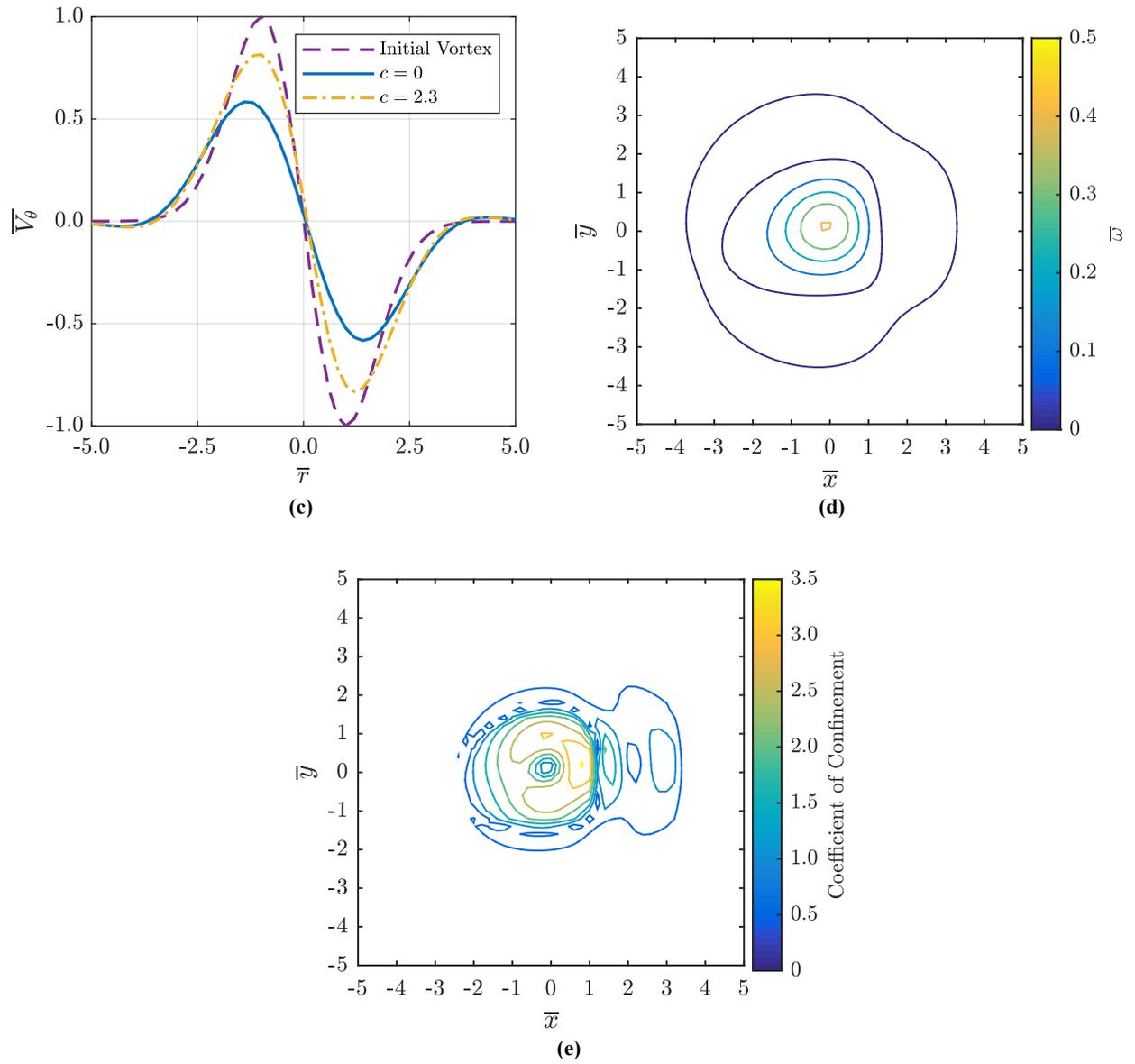

**Fig. 2** Second order spatial discretization using modified VC with c=2.3: (a) kinetic energy (Eq. (9)) normalized by the initial value; (b) absolute value of x-momentum (Eq. (10)) normalized by the initial value (c) tangential velocity, initial vortex and convection results; (d) contours of vorticity at t=0.2 ($\bar{t} = 0.25$); (e) contours of ε value.

## B. Use of TVD Schemes in Combination with Vorticity Confinement

While the pure second-order scheme was properly confined in the previous section with a moderate value for the constant coefficient of confinement parameter, $c$, it could be easily over-confined if the proper value was not selected. This leads to a non-physical solution where the vortex became stronger as time increases. Over-confinement may even lead to the failure of the simulation altogether. To increase the fidelity of the calculations, the van Albada and minmod TVD limiters [20] were applied to the fluxes.

TVD schemes work by reducing second-order schemes to first-order schemes in the presence of oscillations. These oscillations cause values of the limiter function, $\Phi$, to tend to zero in Eq. (11). The second-order scheme was created through linear interpolation of the flux terms shown in Eq. (11). For a pure second-order scheme, $\Phi = 1$, while $\Phi = 0$ for a first order scheme. For TVD implementation in this investigation, the limiter function was either the minmod, Eq. (12), or van Albada, Eq. (13), limiter.

$$F_{i+\frac{1}{2}} = F_i^+ + 0.5\Phi_{i+\frac{1}{2}}^+(F_i^+ - F_{i-1}^+) + F_{i+1}^- + 0.5\Phi_{i+1/2}^-(F_{i+1}^- - F_{i+2}^-) \tag{11}$$

where $\Phi_{i+1/2}^+ = f((F_{i+1}^+ - F_i^+)/(F_i^+ - F_{i-1}^+))$

$$f(x) = \max(0, \min(1, x)) \tag{12}$$

$$f(x) = (x^2 + x)/(x^2 + 1) \tag{13}$$

The same parameters as those used in Sec. IIA were used for numerical computations in this section with the exception of the cross-flow, which was zero for these simulations. The domain size was equal to 20 in the $x$ and $y$ directions.

The results shown in Fig. 3a had VC active with a coefficient of confinement of $c = 2.5$. The velocity profile was normalized by the peak velocity of the initial velocity profile, Eq. (8). These results show that this value of $c$ maintained the vortex well for the pure second-order scheme. However, the computations with the van Albada and minmod limiters were under-confined as the TVD schemes reduced the order of numerical approximation and introduced more numerical dissipation.

The results shown in Fig. 3b utilized VC with $c = 4.0$. For this case, the purely second order simulation experienced significant over-confinement as the maximum velocity of the vortex doubled compared to the initial profile. Additionally, the velocity gradient in the vortex core increased significantly. The first-order simulation is shown to be considerably under-confined with much more numerical dissipation than any of the second-order schemes. The TVD results were closer to the initial vortex profile with the van Albada limiter producing more accurate results compared to the minmod limiter.

Figure 3c is the results of the case with $c = 5.0$. Here, the second order scheme failed completely and produced NaN (not a number) values due to over-confinement. The results of the van Albada differentiable limiter by Eq. (13) also showed signs of over-confinement (compare Fig. 3c to Fig. 3b). The minmod limiter (Eq. (12)) was less affected by the selection of the value of $c$, and the results were similar to those of the simulation using $c = 4.0$. This indicates that the non-differentiable minmod limiter is more robust with respect to a selected value of $c$ compared to the differentiable van Albada limiter. Thus, a larger range of $c$ values would produce suitable confinement of the vortex when using the minmod limiter. However, the level of numerical dissipation using minmod was higher than that of the differentiable limiter which leads to smearing of the vortex without a stronger VC source term.

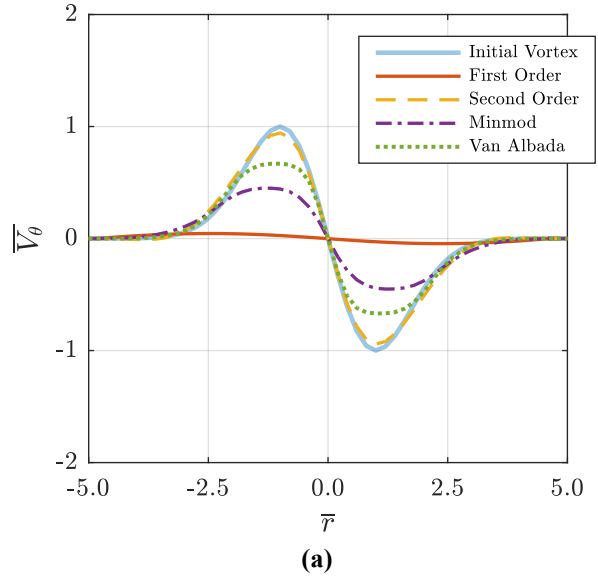

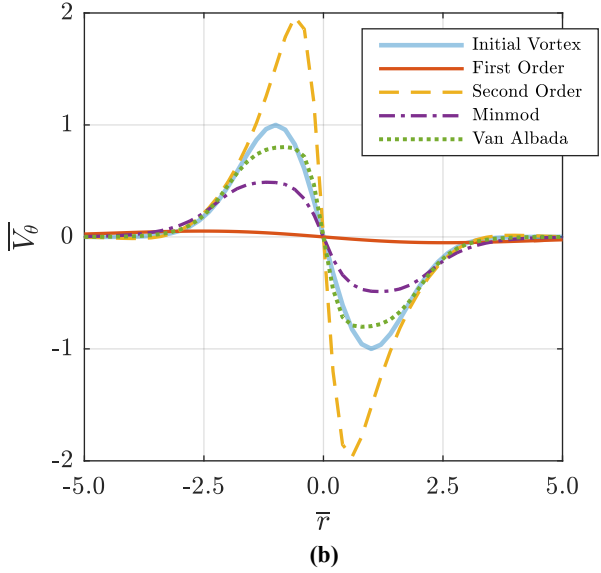
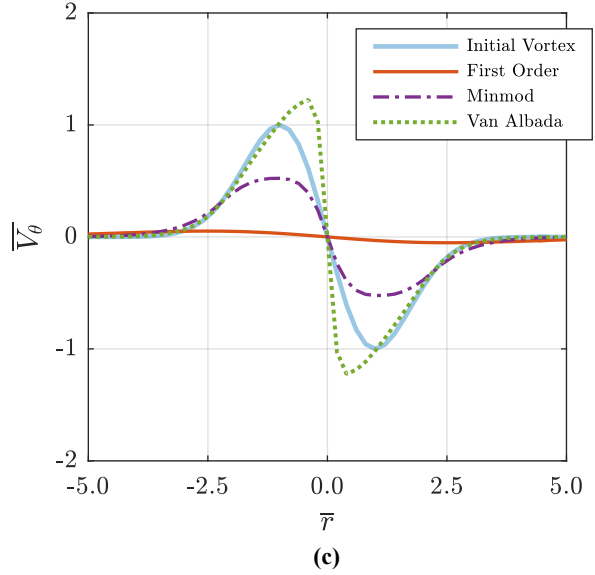

**Fig. 3 Comparison of TVD Schemes applied together with vorticity confinement, (a) $c = 2.5$, (b) $c = 4.0$, and (c) $c = 5.0$.**

### C. Scaling of Numerical Scheme with the Grid Size

One of the main advantages of using the modified VC formulation is that the strength of the source term scales with grid density for order of error for second order schemes, $O(h^2)$. This is a useful property that allows a constant confinement coefficient to be used for various grid densities. In modified VC, the source term scales by Eq. (5) and automatically adjusts its magnitude when all terms other than the leading truncation error are negligible.

Numerical simulations were carried out to determine when the second-order upwind scheme developed for this research has accuracy that is dominated by $O(h^2)$. The simulation domain was 10 units in both the $x$ and $y$ directions. The properties of the initial vortex for these simulations were chosen as $r_0 = 1$ and $\Gamma = 50$, the same as in Sections IIA and IIB, and there was no crossflow so the size of domain in the $x$ direction is smaller compared to prior case. The simulations were carried out for various grid steps ranging from 0.5 to 0.0133. For each case the grid steps are the same in the $x$ and $y$ directions, $\Delta x = \Delta y$. The normalized grid step value $\overline{\Delta x} = \frac{\Delta x}{r_0}$. The accuracy of the scheme was determined through investigation of the maximum numerical value of velocity after a simulation time of 0.1. The value of maximum velocity was then used to determine a value for $U_{diff}$, a nondimensional measure of the reduction in velocity due to numerical dissipation. $U_{diff}$ was defined as $U_{diff} = \frac{|U_0 - U|}{U_0}$ where $U_0$ was the maximum velocity of the initial flowfield and $U$ was the maximum value after the simulation was completed. The computed values of $U_{diff}$ with and without VC for the range of grid steps are shown in Table in Appendix. The absolute value of the difference was used to allow the plotting of over-confinement (larger final velocity compared to initial velocity) using a logarithmic scale. Note that in absence of numerical error $U = U_0$ and $U_{diff} = 0$.

Figure 4 illustrates the results of the grid convergence study. Figure 4a is a plot of $\log_{10}(U_{diff})$ and $\log_{10}(\overline{\Delta x})$ for a set of cases without VC applied and for cases with VC applied where the coefficient of confinement was set to $c = 0.8$ and $c = 0.3$. When the scheme reaches second order accuracy, the plot becomes a linear line for the results without VC. This is an indication that the error is dominated by the first term in the truncation error. Investigation of Fig. 4a shows that the scheme reached second order accuracy at approximately $\overline{\Delta x} = 0.1$. Above this value the results clearly deviate from second order accuracy, increasingly so as the grid coarsens.

The coefficient of confinement parameter was determined to be $c = 0.8$ to achieve minimum error for a grid with $\overline{\Delta x} = 0.1$. This value of $c$ was then applied to other grid densities to check that the confinement parameter does indeed scale properly. The results show that the strength of the modified VC source term scales well if the grid step was small enough to cause the higher order terms in the truncation error to be negligible. As seen in Fig. 4b, $\overline{\Delta x} = 0.1$ and $\overline{\Delta x} = 0.05$ were both well confined (overlap each other and the initial velocity profile) for a confinement coefficient of $c = 0.8$ while the lower density grids exhibit increasing amounts of decay of peak velocity as the grid coarsened. However, grid refinement levels finer than $\overline{\Delta x} = 0.05$ resulted in slight over-confinement for $c = 0.8$. A second set of cases was run for $c = 0.3$, which did not result in over-confinement for any cases in the range of grid steps investigated. In addition to producing no over-confinement, the selection of $c = 0.3$ also resulted in a constant slope in the log-log plot of $U_{diff}$ (Fig. 4a) for grid steps within the interval $0.0167 \leq \overline{\Delta x} \leq 0.1$. The linearity in the log-log plot indicates that for this selection of $c$, the vorticity confinement source term is scaling nearly perfectly with the second-order error of the numerical scheme. The reduction in the numerical error achieved using VC can be evaluated using the slope of $U_{diff}$ in Fig. 4a. For the grid steps interval $0.0167 \leq \overline{\Delta x} \leq 0.1$, the slope of $U_{diff}$ (determined using linear regression) on the log-log plot is 2.865 for the cases without VC, while it is 3.894 for the cases with VC ($c = 0.3$). By using modified VC with an optimal value of $c$ in Eq. (5), the second-order component of the error is nearly cancelled and the order of error approaches fourth order. The non-linearity of the log-log plot for very dense gridding, $\overline{\Delta x} < 0.0167$, shows that the leading error term is not fully cancelled, nevertheless, even for the most refined grid, $\overline{\Delta x} = 0.01$, the error drops more than 3 times compared to the scheme without VC (see Table in Appendix).

For full cancellation of the leading error term in the modified differential equation for second-order upwind schemes [20], the coefficient of confinement, $c$, should be non-constant in space and depend on the third derivative of solving variable, $u_{xxx}$, which is unknown for a general case. Investigation of the VC methodology with the non-constant $c$ will be conducted in the future research.

Note that the larger values of $c$ used in the prior section were selected for $\overline{\Delta x} = 0.2$, for which the second-order of accuracy is not reached. In Fig. 4a, the results with VC further demonstrated its benefit as the numerical error represented by $U_{diff}$ was reduced for the most refined case by nearly two orders of magnitude compared to the simulation without VC. Note that for original VC formulation with constant $\varepsilon$, the scaling of the VC source term with truncation error would not be possible, and $\varepsilon$ required to maintain the original vortex would be a different value for every calculation.

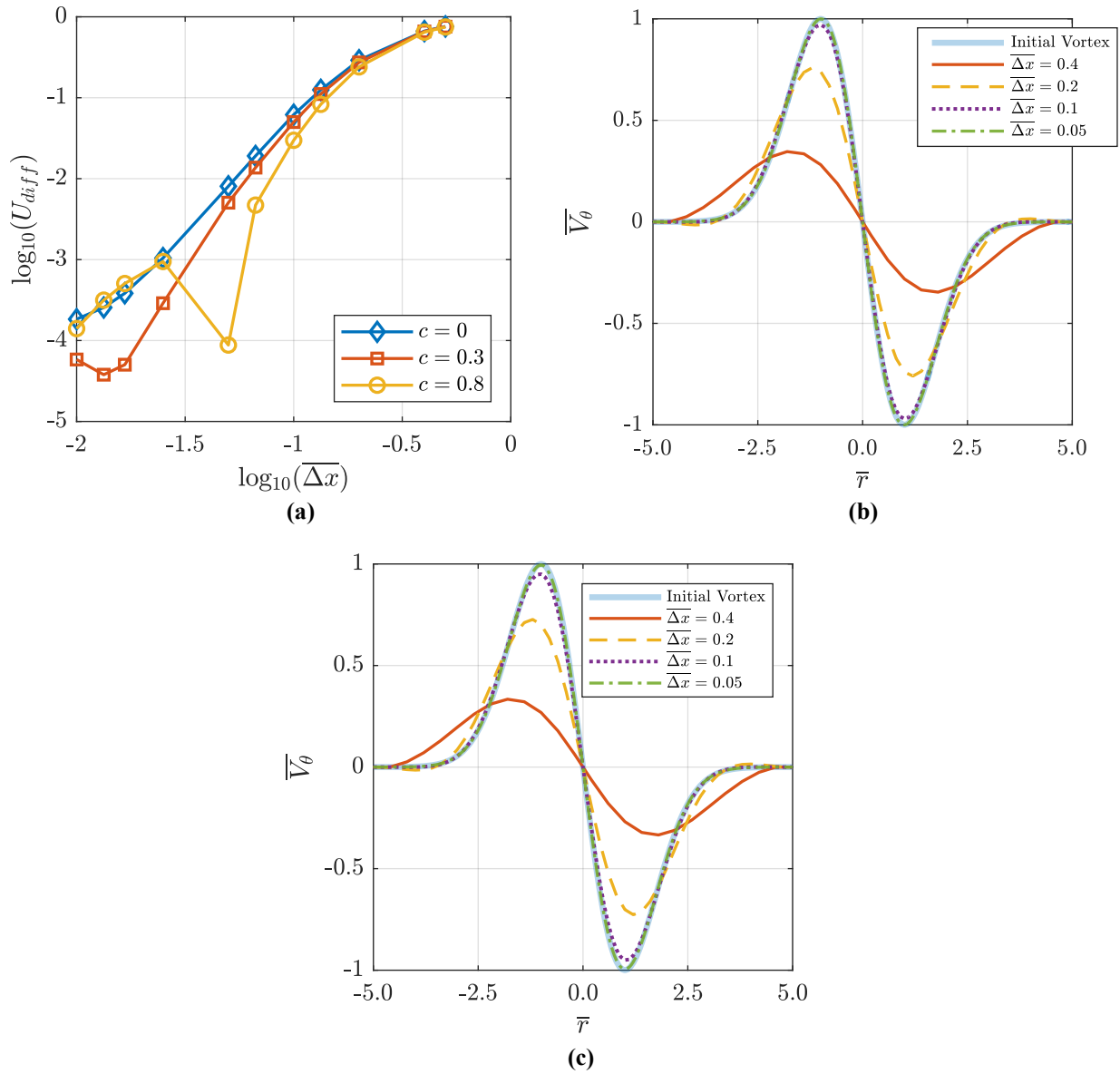

**Fig. 4 Truncation error for various grid step sizes: (a) numerical error as a function of grid step; (b) vorticity confinement balancing the truncation error ($c = 0.8$); (c) vorticity confinement balancing the truncation error ($c = 0.3$)**

For 3-D simulations of flow around wing, it becomes impractical to use a uniform highly refined grid throughout the domain. Therefore, there is no guarantee that those types of simulations will have the same second-order behavior as the well-refined uniform grid simulations conducted in this section. Nevertheless, the scaling of VC with grid size allows modified VC to adapt to various grid cell sizes used for the simulations in the next sections.

**D. Numerical Dissipation of 2D ANSYS FLUENT Simulation**

The appropriate selection of the coefficient of confinement parameter, $c$, is highly dependent upon the level of numerical dissipation which occurs in the simulation. The commercial CFD software ANSYS FLUENT, used for 3D simulation in Sections III and IV, allows the user to select different solvers and solution methods, both of which may affect the amount of numerical dissipation. Additionally, since FLUENT source code is proprietary, some of the algorithms used to increase both robustness and accuracy are not known to the public. Per FLUENT manual, they use multi-stage Runge-Kutta scheme with the non-determined number of stages and, possibly, sub-iterations for reducing errors caused by the non-linearity of problems [27]. In order to understand the level of numerical dissipation in a FLUENT simulation compared to the authors' MATLAB code, simulations were conducted in which the solution methods selected in FLUENT were as similar as possible to the MATLAB code. The simulated vortex and computational domain were the same as the one used in Section II C. The grid step was selected to be $\overline{\Delta x} = \overline{\Delta y} = 0.05$. The simulations were conducted only to analyze the level of numerical dissipation without VC.

The solver type which matched most closely with the MATLAB code was the density-based method using a transient time formulation with explicit time stepping. The selection of Steger-Warming flux splitting, used in the MATLAB code, was not available in FLUENT; therefore, the selection of Roe-FDS flux splitting was used for the comparison while another flux method used in FLUENT [27], Advection Upstream Splitting Method (AUSM), has the approximately doubled value of error compared to the Roe flux. The details of the simulation settings used in Fluent are summarized in Table 1.

Table 1  FLUENT model settings, solver settings, and solution methods for comparison to MATLAB code.

| FLUENT Settings | |
|---|---|
| **Solver and Models** | |
| **Solver** | Density-based |
| **Viscous Model** | Inviscid |
| **Density** | Ideal Gas, Air |
| **Energy Equation** | On |
| **Solution Methods** | |
| **Transient Formulation** | Explicit |
| **Flux Type** | Roe-FDS |
| **Flow** | Second Order Upwind |

The results of the FLUENT and MATLAB simulations showed that the simulation conducted using FLUENT has significantly less numerical dissipation compared to the simulation conducted using the MATLAB code. This is shown in Fig. 5, where the velocity profile from the result of the FLUENT simulation experienced an 8.2% decrease of the maximum velocity compared to the initial vortex velocity. For the simulation conducted using the MATLAB code, the decrease in maximum velocity from the initial velocity profile was 33.8%.

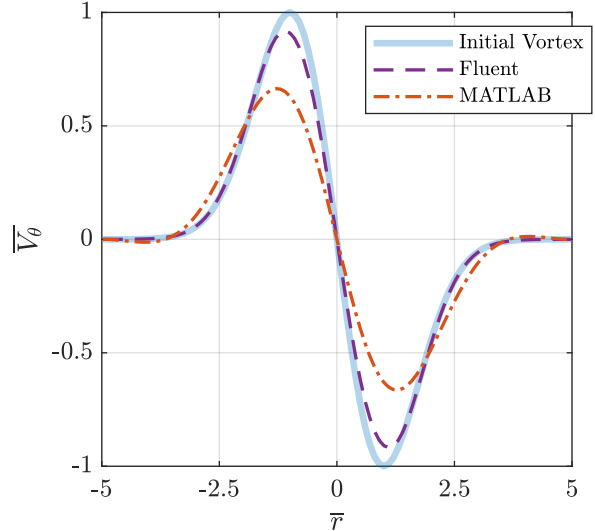

**Fig. 5 Velocity profile comparison for Fluent and MATLAB simulations of a stationary vortex at $t = 1$**

In Section III, one may notice that the value of $c$ for the 3D simulation of wingtip vortices using FLUENT ranged from 0.04 to 0.075, which was approximately two orders of magnitude lower than the value of $c$ used in Section II B. The lower value of $c$ required for 3D simulation in FLUENT indicates that these simulations had less numerical dissipation compared to the MATLAB code, which is consistent with the result presented in Fig. 5.

### III. Application of VC to Induced Wingtip Vortices

**A. Computational Setup**

The external flow of air over a 3-D NACA0012 wing was modeled using ANSYS FLUENT software [27] for the determination of induced drag for angles of attack (AoAs) of 4, 6 and 10 degrees and Mach numbers of 0.3, 0.5, and 0.6. The ANSYS FLUENT solver settings selected were a pressure-based solver which applied pressure-velocity coupling algorithm (SIMPLE) [26] along with the energy equation and second-order upwind spatial discretization for convective terms in the Euler equation. The pressure-based solvers are based on a formulation in which the momentum equations are solved for the velocity components using the pressure field from prior iteration. The pressure is then recalculated using the continuity equation. When the pressure changes, the velocity field is updated too. The sequence is repeated iteratively until the flowfield satisfies continuity, momentum and energy equations [20].

The VC body force term was added to the core ANSYS Fluent solver by development of a user-defined function (UDF). The addition of VC was found to increase the computation time per iteration by 7.6% and to increase the number of iterations required for convergence by 42%. Convergence was achieved when normalized residuals, determined by ANSYS Fluent, fell below $10^{-5}$. A NACA0012 airfoil profile was used to construct a 3D wing with an aspect ratio of 6.67. The computational domain (Fig. 6) consisted of 1.38 million cells and extended 14 chord lengths away from all surfaces of the wing. This setting resulted in grid cells which had an average step size of 0.025C (0.025 x chord length) on the surface of the wing. The mesh around the surface of the wing is shown in Fig.7a. The region of refinement was designated from the tips of the wing to the edge of the computational domain oriented along the vortex path as seen in Fig. 6. This resulted in a grid step size of approximately 0.06C. The computational grid in the vortex advection region is illustrated in Fig. 7b. A grid convergence study is included in the results presented in the next subsection.

Both pressure- and density-based solvers were investigated to find the method with better convergence properties [20,27]. The converged computational results were nearly identical; however, the density-based solver did not converge as quickly and was more oscillative in terms of the behavior of the residuals. Consequently, the pressure-based solver was used in subsequent cases and in all cases presented in this paper. FLUENT model settings for inviscid computations are summarized in Table 2.

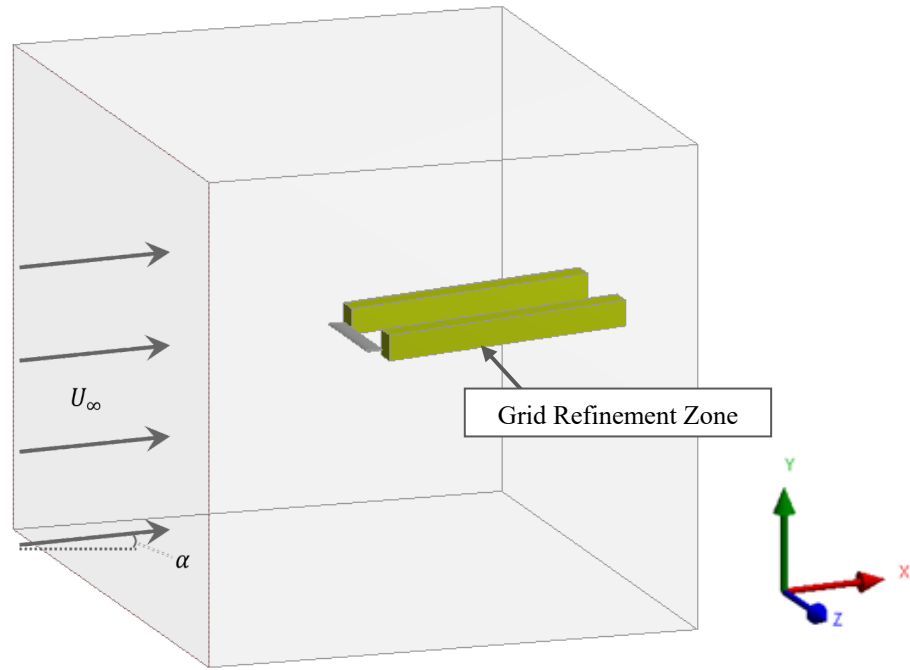

**Fig. 6 Computational domain for 3D wing**

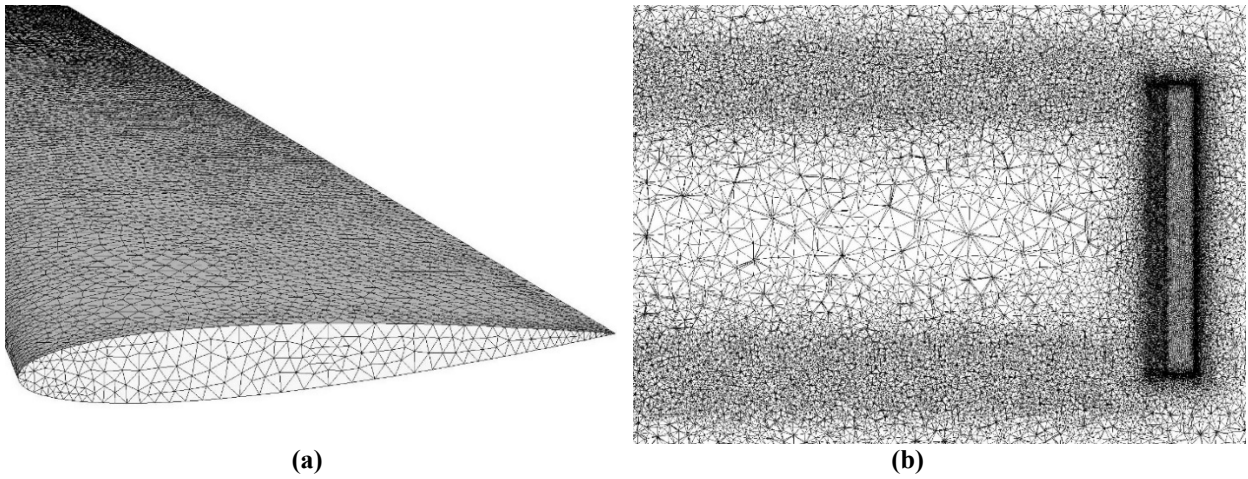

**(a)**                                            **(b)**
**Fig. 7 Discretized computational domain; (a) surface of the wing; (b) wing and vortex advection region.**

Table 2  Model settings, solver settings, and solution methods for inviscid wing simulations

| FLUENT Settings | |
|---|---|
| **Solver and Models** | |
| **Solver** | Pressure -based |
| **Viscous Model** | Inviscid |
| **Density** | Ideal Gas, Air |
| **Energy Equation** | On |
| **Solution Methods** | |
| **Pressure-Velocity Coupling** | SIMPLE |
| **Density, Momentum, Energy** | Second Order Upwind |

**B. CFD Results for Induced Drag Obtained by Integration in Trefftz Plane**

In order to determine the accuracy of the proposed numerical approach, the results were compared to those obtained by inviscid lifting line theory. Analytical lifting line theory, developed to approximate lift exerted on a 3-D wing, can also be used to approximate the induced drag force coefficient, $C_{D,i}$.

$$C_{D,i} = \frac{C_L^2}{\pi e_{sp} AR} \tag{14}$$

In Eq. (14), $C_L$ is lift coefficient computed by classical inviscid aerodynamic theory [29] , $e_{sp}$ is the span efficiency factor which accounts for non-elliptic circulation distribution and $AR$ is the aspect ratio.

The inviscid CFD computations involve the same assumptions used in the lifting line theory. The Prandtl–Glauert [28] compressibility correction is used to determine theoretical induced drag for Mach numbers 0.5 and 0.6. Since only subsonic flows were investigated, AoAs of 10 and 6 degrees were not evaluated at $M = 0.6$ and AoA of 10 degrees was not evaluated at $M \geq 0.5$ as flow is transonic for these cases. A span efficiency factor of 0.94 was used in these calculations of induced drag to account for the non-elliptic circulation distribution [29]. More details of calculation of induced drag are provided in Thesis of the first author [25, p. 34].

The goal of use of VC was to maintain the strength of the tip vortex after its formation. Thus, in order to reduce interference with vortex roll-up, the VC body force terms were not activated until the trailing vortices have convected 0.25 Chords (C) downstream of the trailing edge [2,8].

As seen in Fig. 8, without application of VC the wake integral method provides different results for drag coefficient, $C_D$, when integration is conducted at different Trefftz plane locations. For each case depicted in Fig. 8, representative values of $c$ are shown corresponding to under-confinement, over-confinement and optimal confinement situations. When VC is used to counteract numerical dissipation and an optimal confinement parameter is selected, the results become approximately independent of the Trefftz plane location. Lifting line theory was used as the basis of comparison to determine the accuracy of both the near-field and far-field techniques.

Table 3 shows that surface integration consistently over-predicted the drag force, sometimes causing the drag coefficient to be over two times the value obtained by lifting line theory. The wake-integral method consistently under-predicted the drag force but with significantly less error in comparison to surface integration. The drag coefficient was typically under-predicted by 25%, with higher accuracy at lower Mach numbers and lower accuracy at higher Mach numbers. Lifting line theory generally does not include effects such as the non-uniform downwash of an elliptically loaded wing and the non-planar character of the wake shed from a curved trailing edge [30,31]. These effects become more pronounced at larger Mach numbers that contributes to the discrepancy between far-field integration results and those obtained by lifting line theory.

Although the formulation of VC (Eqs. 3-7) was designed to account for vortex strength and grid density, it was still necessary to tune the confinement parameter for individual cases. To select the value of confinement parameter, $c$, for a particular case presented in Table 4, the drag coefficient, obtained by integration in the Trefftz plane, should be invariant with the location of the Trefftz plane so as both under-confinement (drag decreases with the distance from the wing to the Trefftz plane) and over-confinement (drag increases with the distance from the wing to Trefftz plane) are avoided. The optimal confinement parameter was relatively consistent for different Mach numbers of the same AoA as seen in Table 4. For the values of $c$ presented in Table 4, the drag coefficients in Table 3 were obtained.

As the AoA becomes larger, the coefficient of confinement parameter, $c$, became nearly constant. This consistency allowed for a faster and more accurate tuning for the proper confinement parameter. Even when the confinement parameter was not large enough to completely counteract numerical dissipation, the results were still notably better than those without VC implemented.

The VC technique was most useful at the AoAs presented in this section (4 degrees and above) as the amount of numerical dissipation that exists in these cases is substantially larger. For lower AoAs, induced drag is not as significant as profile drag, though still not negligible. Per presented computations for NACA0012 wing, a reasonably conservative approach could use $c = 0.0475$ for all simulations for AoA lower than 6 degrees and $c = 0.0375$ for all AoAs between 6 and 10 degrees. Though this piecewise-constant approximation of parameter $c$ would not completely eliminate numerical dissipation, it would make certain that there is no violation of physics provided by over-confinement and would still be a significant improvement over no confinement.

One may notice that the value for $c$ was lower, by nearly two orders of magnitude, for FLUENT simulations when compared to MATLAB simulations in Section II. Numerical experiments were carried out to determine the reason for this difference. A stationary 2D Taylor vortex, with the same properties as in Section II B, was simulated in FLUENT using its transient, density based solver. The resulting $c$ value for the 2D FLUENT computation were on the same order as those from the 3D wingtip vortex. It was therefore determined that FLUENT, perhaps through the use of some proprietary schemes (see Section II D), consistently exhibits lower values of numerical dissipation compared to the MATLAB code used in Section II.

**Table 3  Drag coefficients obtained by surface integration and wake-integral method compared to lifting line theory**

| Drag Coefficient | | | | | |
| --- | --- | --- | --- | --- | --- |
| M=0.3 | | | | | |
| AoA | Lifting Line Theory | Surface Int. | Wake-Int. with VC | Surf. Int. Error % | Wake-Int. Error % |
| 4 | 0.00544 | 0.0116 | 0.0043 | 113 | -21 |
| 6 | 0.0122 | 0.021 | 0.0104 | 72 | -15 |
| 10 | 0.034 | 0.052 | 0.027 | 53 | -21 |
| M=0.5 | | | | | |
| 4 | 0.0072 | 0.0131 | 0.0055 | 81 | -24 |
| 6 | 0.016 | 0.0238 | 0.0121 | 49 | -24 |
| M=0.6 | | | | | |
| 4 | 0.0084 | 0.0143 | 0.0061 | 70 | -27 |

**Table 4 Tuned confinement parameters**

| Confinement Parameters | | | |
| --- | --- | --- | --- |
| AoA | M=0.3 | M=0.5 | M=0.6 |
| 4 | 0.075 | 0.07 | 0.065 |
| 6 | 0.0525 | 0.05 | |
| 10 | 0.04 | | |

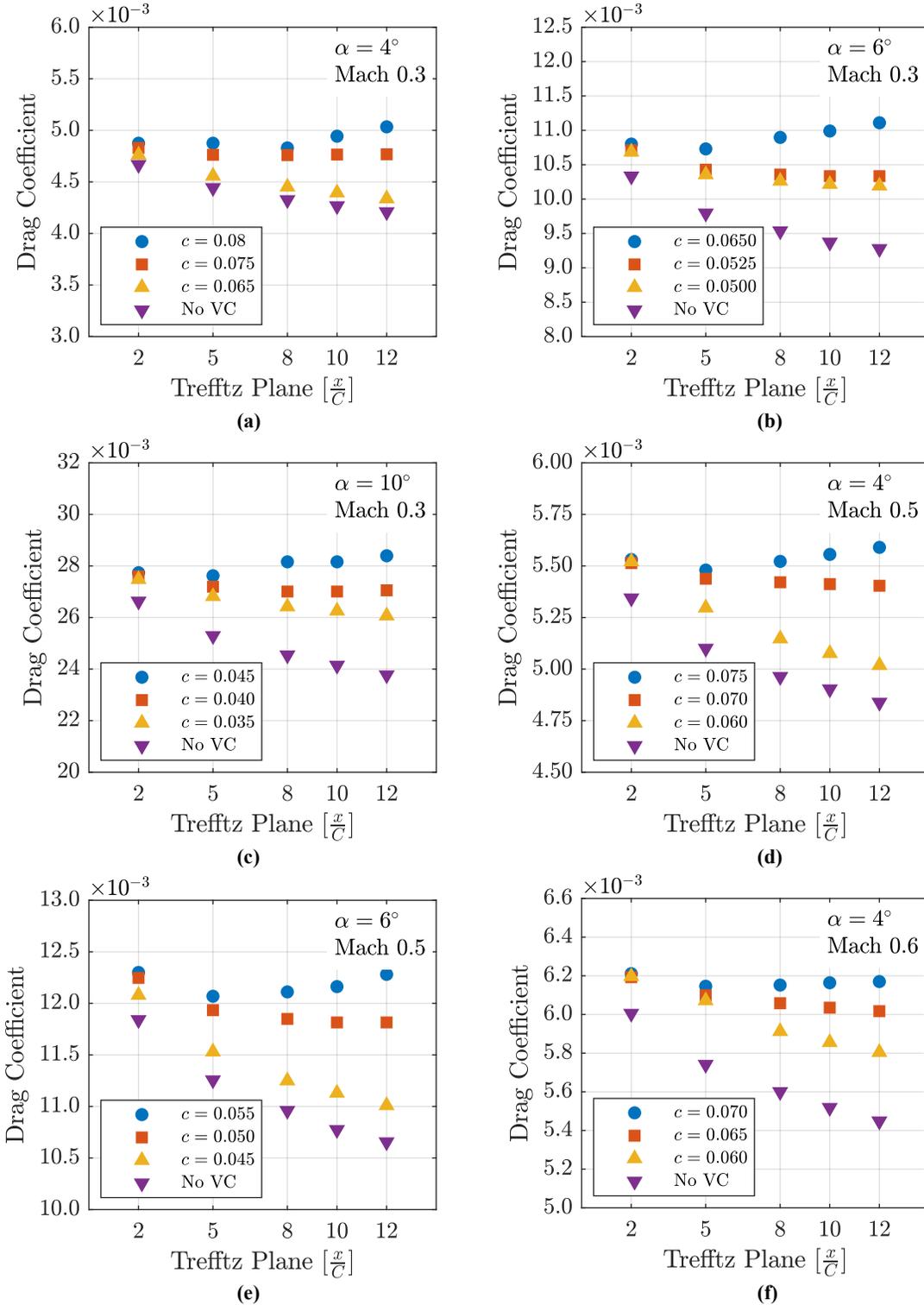

**Fig. 8 Induced drag coefficient for inviscid simulations: (a), AoA = 4° and M=0.3; (b), AoA = 6° and M=0.3; (c), AoA = 10° and M=0.3; (d), AoA = 4° and M=0.5; (e), AoA = 6° and M=0.5; (f), AoA = 4° and M=0.6.**

To evaluate the effects of grid density on the results, three additional meshes were created by reducing the element size for both the face sizing on the surface of the wing and the grid step in the vortex advection region by approximately one third. Additional simulations were carried out using these refined meshes for the case of AoA = 4° and M=0.3. These new meshes are denoted as refined meshes A, B and C. The finite-volume linear size and resulting number of grid cells are listed below in Table 5. Drag coefficients for the simulations were again evaluated using surface integration alongside wake integration with VC.

**Table 5: Grids used for convergence study**

| Grid | Wing Face Sizing (chords) | Grid Sizing in Refinement Region in Figure 6 (chords) | Total Grid Cells (million) |
|---|---|---|---|
| Standard | 0.025 | 0.06 | 1.38 |
| Refined Mesh A | 0.0167 | 0.04 | 2.48 |
| Refined Mesh B | 0.011 | 0.0267 | 5.96 |
| Refined Mesh C | 0.007 | 0.0206 | 13.41 |

The number of grid cells for modeling of airflow around wing are comparable to those used in literature. For example, the recent study [32] computes viscous airflow with detailed resolution of boundary layer and with the wing aspect ratio equal to 12. They found that a grid sizing of 5–9 million tetrahedral cells is adequate for a rectangular wing. The typical grids for a general wing averaging between 9–12 million tetrahedral cells. Refs. [1,4] considered meshes from 5M to 30 M. However, these meshes are intended for viscous/turbulent airflow around the full model of airplane as opposed to inviscid flow about an isolated wing.

The results of the grid convergence study are presented in Table 6 in terms of induced drag coefficient. The refinement of the mesh shows a trend of increasing drag coefficient for wake integration and decreasing drag coefficient for surface integration. For induced drag evaluation by surface integration, the drag coefficient of the most refined mesh C is approximately doubled compared to that of lifting line theory (see Table 3). For induced drag evaluation by wake integration, the drag coefficient of refined mesh C is only 10% lower than that obtained by the lifting line theory. The change in the induced drag obtained by the wake integration between the standard mesh and refined mesh A is greater than that between refined mesh A and refined mesh B while the results obtained on grids B and C are close to each other.

This indicates that the drag coefficient is asymptotically approaching a grid independent result. The near-field drag is grid-dependent, and the extrapolated lifting line value is not achieved using grids typical for practical computations. Wake integration coupled with vorticity confinement remains more accurate and closer to lifting line theory.

Table 6: Result of grid convergence study in terms of drag coefficient for AoA = 4° and M=0.3

| Grid | Drag Coefficient | |
|---|---|---|
| | Wake Integration | Surface Integration |
| Standard | 0.0043 | 0.0116 |
| Refined Mesh A | 0.0047 | 0.0107 |
| Refined Mesh B | 0.0049 | 0.0104 |
| Refined Mesh C | 0.0050 | 0.0096 |

**C. Role of TVD Limiter**

An investigation into the appropriate choice of TVD limiter and its effect on vortex dissipation in the frame of the VC-TVD approach is presented in this section. Two different TVD limiters (differentiable and minmod) were investigated through application to 3-D simulations of vortex shedding. The 3-D simulation results showed consistent behavior with the 2-D simulation results of Sec. II B.

In an extension of the investigation conducted in the previous section, a simulation utilizing the differentiable limiter available in ANSYS FLUENT was conducted to determine the coefficient of confinement. The differentiable limiter used in ANSYS FLUENT is a modified form [33] of a limiter which was originally proposed by Venkatakrishnan [34]. Figure 9 shows the comparison of results for both the differentiable limiter and the minmod limiter without VC applied. As expected, the differentiable limiter had a lower level of numerical dissipation as the level of vorticity reduced at a lower rate than that using the minmod limiter.

The 4 degree AoA and Mach 0.3 case was investigated previously for the minmod TVD limiter. The coefficient of confinement for that case was determined to be $c = 0.075$ [2,8]. Due to the results in Sec. II B, it was expected that the differentiable limiter would produce less numerical dissipation and would therefore require a lower value for $c$. For the plots of tip vortex with VC applied in Figs 9 and 10, vorticity is normalized by $\bar{\omega} = \omega \frac{t_{wing}}{U_\infty}$ where $t_{wing}$ is the wing thickness ($t_{wing}=0.12C$ for NACA0012) and $U_\infty$ is the freestream velocity; the position is normalized by $\bar{z} = \frac{z}{t_{wing}}$ and $\bar{y} = \frac{y}{t_{wing}}$ with $y$ and $z$ equal to zero at the edge of the wing.

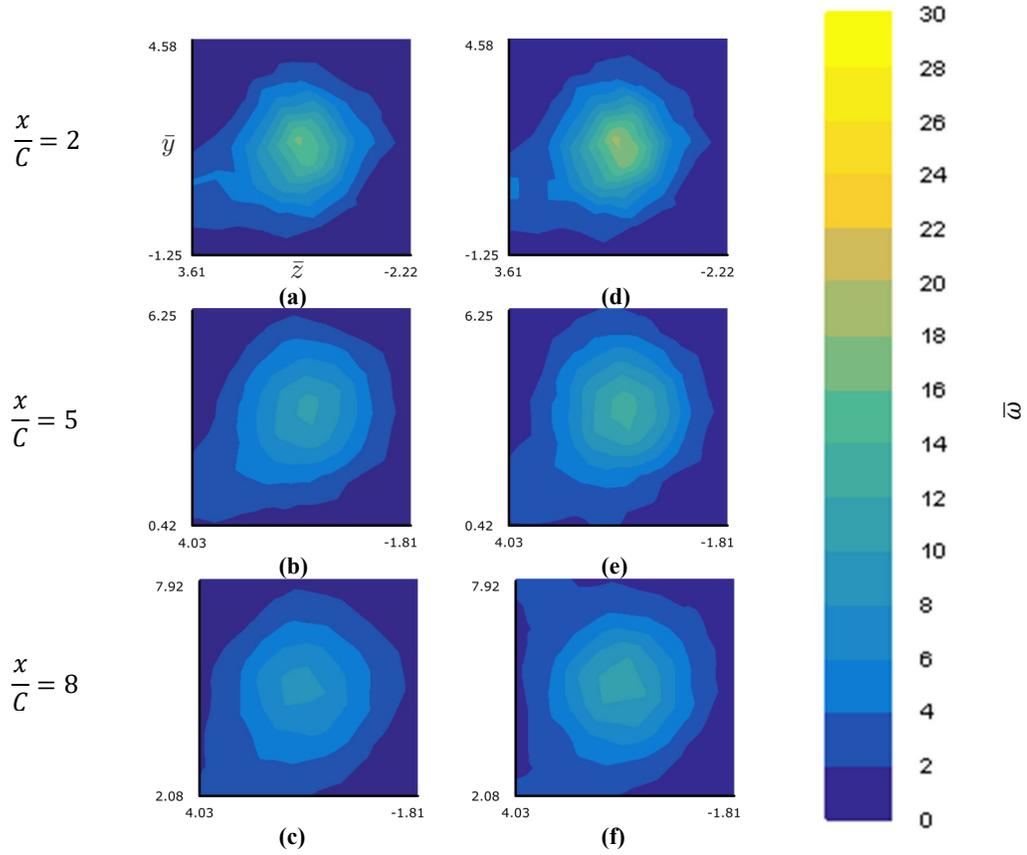

**Fig. 9** Vorticity in tip vortices obtained without VC enabled: (a-c) minmod limiter; (d-f) differentiable limiter; (a,d) at distance of 2 Chords (2C) from trailing edge; (b,e) at distance of 5C from trailing edge; (c,f) at distance of 8C from trailing edge. Scale for contour plots is shown on the right side.

Since the level of dissipation was less for the differentiable limiter, it was expected that the $c$ value for proper confinement would be lower than that for the minmod limiter. It was found that $c = 0.0525$ is optimal for VC combined with TVD using the differentiable limiter which was 30 percent lower than the vorticity confinement value for the minmod limiter. Figure 10 shows a comparison of the differentiable and minmod limiter with VC enabled. VC preserved the vortex strength equal to that at 2 chord lengths from the wing for both limiters if the corresponding optimal coefficient of confinement was used.

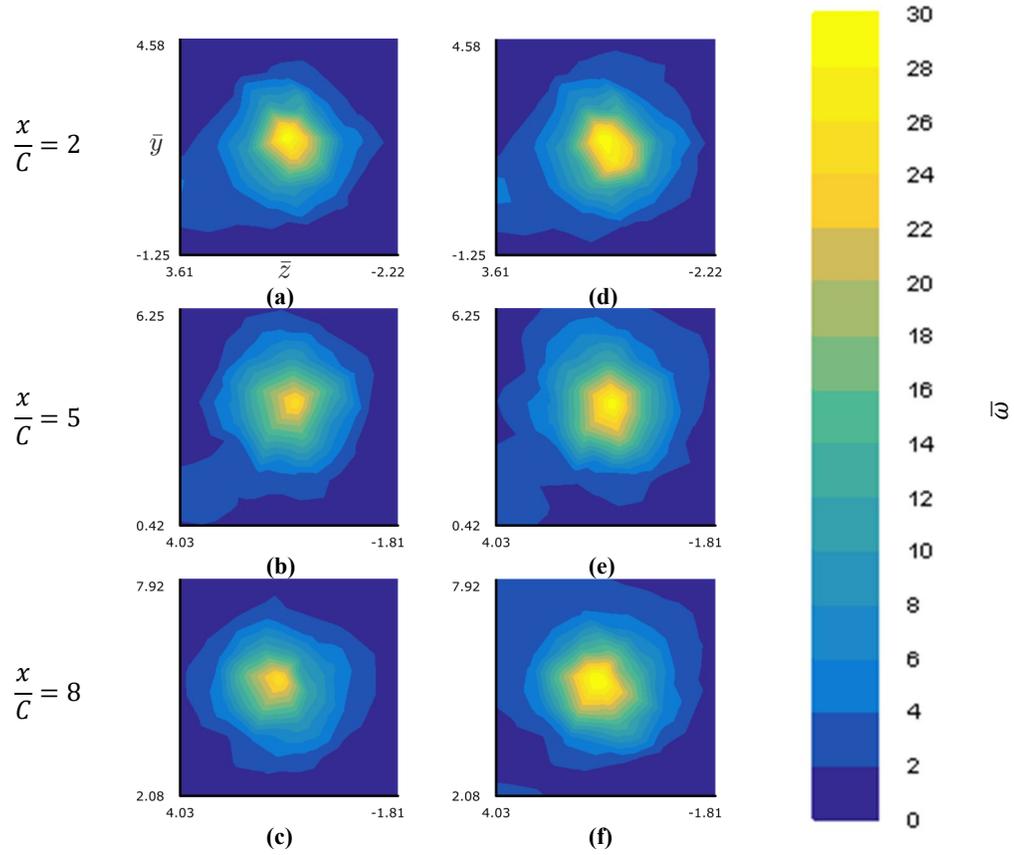

**Fig. 10 Vorticity in tip vortices obtained with VC enabled: (a-c) minmod limiter; (d-f) differentiable limiter; (a,d) at distance of 2C from trailing edge; (b,e) at distance of 5C from trailing edge; (c,f) at distance of 8C from trailing edge.**

## IV. Comparison to Experiments and use of VC with Turbulence Model

In this section, CFD modeling using VC is compared to experimental measurements of swirl velocity in tip vortices. To accurately recreate the experiment numerically, VC was applied to CFD modeling of a tip vortex in order to prevent the tip vortex from experiencing nonphysical numerical dissipation while applying turbulence models to simulate physical dissipation. The results presented in this section are a comparison of computational results produced using the proposed VC method to experimentally measured mean vortex velocity profile in a turbulent wake. The appropriate coefficient of confinement was determined by using an initial guess based upon the results in Section III B. As in the previous section, the appropriate coefficient of confinement was determined by running inviscid simulation. Also, recall (see Section III) that the VC body force terms was not activated until the trailing vortices had convected 0.25C downstream of the trailing edge and therefore VC was not applied within the boundary layer.

Experimental wingtip vortex measurements were conducted using a NACA0012 airfoil profile with a rectangular planform and blunt tip in a wind tunnel test by Davenport et al. [35]. The fixed wing had a chord length of 0.203 m and an aspect ratio of 8.66. The angle of attack for the experiment was $\alpha = 5°$ with freestream velocity $U_\infty = 38.1$ m/s.

The geometry and flow conditions were replicated and two meshes were created to show the effect of mesh parameters on the simulation of tip vortices. The Reynolds stress equations turbulence model (RSM) [27] was investigated for accuracy in modeling of vortex evolution. The velocity in Figs. 11 and 12 is normalized by $\overline{V_\theta} = \frac{V_\theta}{U_\infty}$ while the position is normalized by $\bar{r} = \frac{r}{c}$.

An unstructured mesh, referred to herein as mesh I, with the same meshing parameters (though slightly different geometry) as the mesh in Sec. III was created and resulted in 1.5 million cells. It was determined using inviscid simulations that the appropriate coefficient of confinement for mesh I was $c = 0.04$. The turbulent simulation was then conducted using the RSM turbulence model. However, the simulation did not accurately represent peak velocity or velocity gradients as the grid was too coarse. Experiments show that there is a high gradient of velocity that is present in the vortex core and it is necessary to have a more highly refined grid to properly model the vortex peak velocity and core size.

A second mesh, referred to herein as mesh II, was created as a structured C-grid in order to more accurately model vortex velocity. The mesh only models half the wing span and makes use of the symmetry boundary condition in ANSYS FLUENT in order to increase grid density. The mesh setup resulted in 5-6 cells between peak vortex velocities and resulted in the creation of a grid with 10.6 million cells. Note that the second-order property of the scheme was shown to require approximately 20 cells per vortex diameter (see Fig. 4a) which was not reached by using the mesh II despite its large number of grid cells.

For use of VC in mesh II, it was necessary to change the definition of the characteristic length scale from that used in the unstructured mesh (Eq. (7)). Unstructured grid generation results in values for grid step that are nearly isotropic ($\Delta x \cong \Delta y \cong \Delta z$). However, for mesh II the value of $\Delta x$ became far larger than the other two-dimensional sizes. The step $\Delta x$ becomes the dominant term in the Eq. (7) for $h$. It was found, however, that this term had the lowest impact on numerical dissipation of the vortex. The equation for $h$ was therefore modified to Eq. (15).

$$h = \sqrt{(\Delta y)^2 + (\Delta z)^2} \qquad (15)$$

Using this approach, the coefficient of confinement for mesh II was determined to be $c = 0.032$. Using Eq. (15), grid steps in the cross-section in which the vortex rotates were accounted for, while the grid step along the vortex axis was neglected.

Figure 11a shows the results of the RSM turbulence model without VC. The results show a reduction in the velocity gradient. The vortex core spreads slightly when the vortex is convected from 5C to 10C downstream which was not observed in the experimental results and was the result of numerical dissipation. Figure 11b shows the results of the RSM turbulence model combined with VC. A direct comparison of this final method with the experimental values is provided in Fig 12. The combination of VC with the RSM turbulence model shows practically no vortex decay in terms of velocity gradient and peak velocity and is the most consistent with the experimental results.

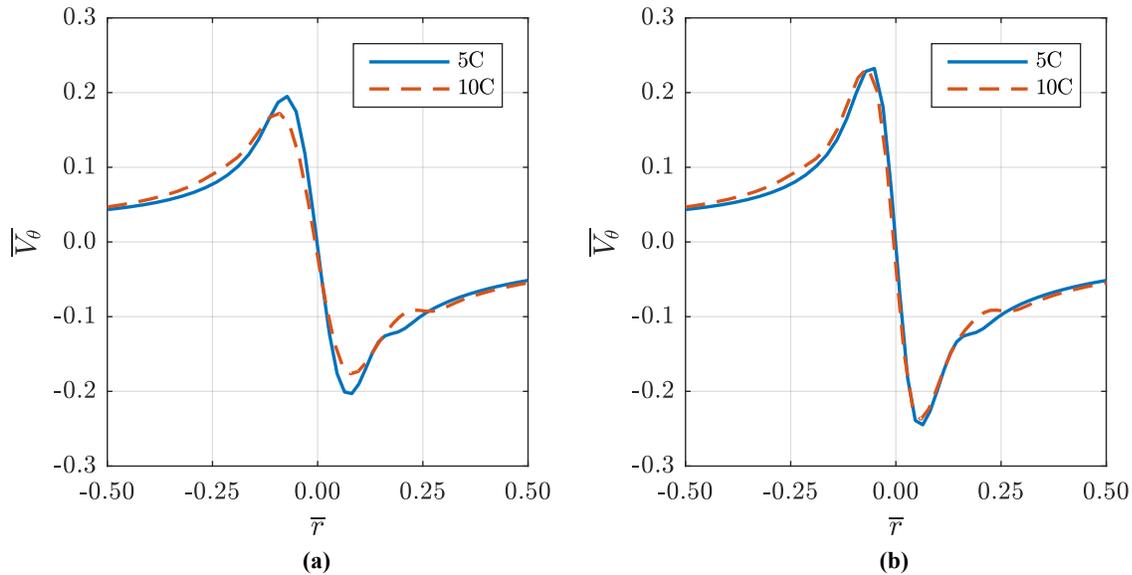

**(a)** **(b)**

**Fig. 11 Results of modeling of tip vortices using Mesh II; (a) RSM turbulence model without VC; (b) RSM turbulence model with VC.**

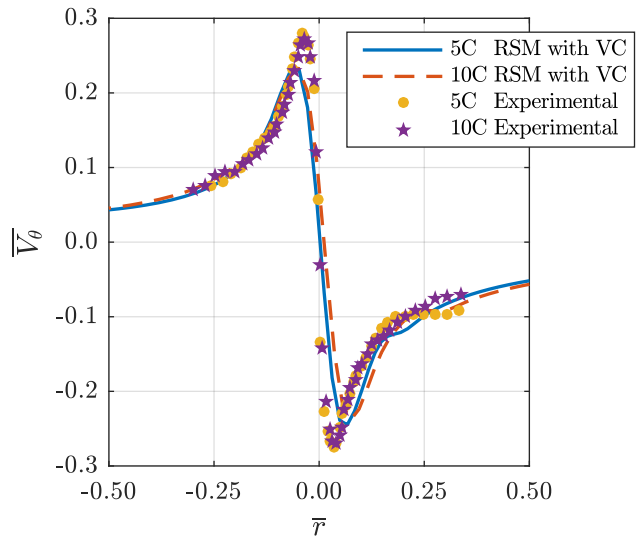

**Fig. 12 Comparison of CFD computations using RSM turbulence model with VC and experimental velocity profile**

## V. Conclusions

It has been shown that surface integration of pressure forces acting on a wing results in the over-prediction of induced drag while Trefftz plane integration results in an under-prediction of drag force which is exacerbated, due to numerical dissipation, as the Trefftz plane location moves further from the wing. Vorticity Confinement (VC) significantly reduces numerical dissipation for convected vortices with modest computational overhead (increased time per iteration by 7.6% and number of iterations by 42% for FLUENT implementation) which makes VC useful for the Trefftz plane-based evaluation of induced drag on wings.

The 2-D simulations of a Taylor vortex using a uniform grid were conducted first to show that VC significantly reduces numerical dissipation. The maximum velocity and vorticity values were significantly improved over the case without VC after the appropriate VC parameters were determined. The 2-D results showed that second-order schemes combined with TVD limiters and VC have a significantly improved accuracy in numerical representation of vortices compared to numerical schemes without these features. The minmod TVD limiter showed a better level of numerical stability and did not experience over-confinement while the van Albada differentiable TVD limiter experienced over-confinement for larger values of confinement coefficient. Nevertheless, differential limiters are advantageous as they introduce smaller amount of numerical dissipation.

Using the combination of TVD with the minmod limiter and VC as described above, the application of VC to far-field evaluation of induced drag in inviscid flow was extended to higher subsonic inviscid Mach number flows (up to Mach 0.6) and larger angles of attack (AoA), up to 10 degrees. The proper confinement parameter within a wide range of AoAs and Mach numbers was determined for each case so that the drag coefficient became independent from the location of the Trefftz plane. This allowed for more accurate elimination of numerical dissipation while retaining the integrity of the simulation by not over-confining. VC was shown to be an effective method to counteract numerical dissipation in far-field integration in determination of induced drag. For an optimal confinement parameter, VC can, to a large extent, negate numerical dissipation.

The VC method was used together with the RSM turbulence model to eliminate numerical dissipation associated with the second-order upwind schemes while retaining physical turbulent dissipation. When combined with VC, the RSM turbulence model accurately simulated the evolution of a wing-tip vortex.

To put succinctly, presented results of optimized VC combined with wake-integral method of drag prediction (Trefftz plane integration) have shown:

1. The wake-integral method was more accurate than surface integration for prediction of the induced drag of 3-D wings. It predicted the induced drag with the value closer to theoretical lifting-line value. VC with a proper value of the confinement coefficient made the wake-integral drag prediction invariant of the Trefftz plane location.

2. The accuracy of wake-integral method was further improved by VC especially for higher angle of attack (up to 10 degrees) and higher Mach number (up to 0.6).

3. VC preserved trailing vortices with modest computational overhead and can be used in combination with TVD limiters for fluxes.

4. VC can be combined with RANS-based computations of turbulent flow using the value of confinement coefficient close to that for inviscid flow. Therefore, the confinement coefficient can be selected by comparison of inviscid CFD results with analytical lifting line theory.

A future direction of research involves investigation of methods for automating VC. As seen in the current study, the VC method proved itself useful if the proper confinement coefficient was utilized. If the process of determining the VC coefficient can be automated, VC may be useful in more broad engineering applications. The method may then, for example, be used in winglet design optimization to reduce the strength of tip vortices. In a broader sense, the future numerical research based on combination of vorticity confinement with entropy cut-off will help to design lift- and thrust- generating air vehicular systems that may include stationary, flapping and rotating wings, and propellers.

## Appendix

Table: Computational error of vortex model with VC (c=0.3 and c=0.8) and without VC (c=0)

|  | $U_{diff}$ | | |
| --- | --- | --- | --- |
| Grid Step | c=0 | c=0.3 | c=0.8 |
| 0.5000 | 7.60E-01 | 7.55E-01 | 7.48E-01 |
| 0.4000 | 6.66E-01 | 6.59E-01 | 6.47E-01 |
| 0.2000 | 2.91E-01 | 2.73E-01 | 2.41E-01 |
| 0.1333 | 1.25E-01 | 1.11E-01 | 8.30E-02 |
| 0.1000 | 6.18E-02 | 5.01E-02 | 2.98E-02 |
| 0.0667 | 1.91E-02 | 1.37E-02 | 4.71E-03 |
| 0.0500 | 8.06E-03 | 5.05E-03 | 8.80E-05 |
| 0.0250 | 1.05E-03 | 2.89E-04 | 9.47E-04 |
| 0.0167 | 3.84E-04 | 5.03E-05 | 5.06E-04 |
| 0.0133 | 2.56E-04 | 3.78E-05 | 3.16E-04 |
| 0.0100 | 1.84E-04 | 5.83E-05 | 1.41E-04 |

**Acknowledgements**

This research was supported by Army Research Office (ARO) grant, NSF I-Corps Sites at the University of Akron, AFRL/Ohio: DAGSI Student and Faculty Grant RB09-8, and ASEE/AFRL summer faculty and student award in 2016.

**List of Figures**

**List of Tables**